\documentclass {aa}

\usepackage{aabib99}

\usepackage{graphicx}

\begin{document}

\title{Distribution of black hole binaries around galaxies}

\titlerunning{Distribution of BH binaries...}

\author{Krzysztof Belczy{\'n}ski, Tomasz Bulik, Wojciech
Zbijewski}

\authorrunning{Belczy{\'n}ski et~al.}
 
\institute{Nicolaus Copernicus Astronomical Center, 
Bartycka 18, 00-716 Warszawa,Poland}

\date{Received ................, Accepted ....................}

\thesaurus{02.07.2, 08.02.03, 08.05.03}
 
\maketitle 
 
\begin{abstract} 
{  Compact object mergers are one of the favorite models
of GRBs. It has been noted that in opposition to the collapsars,
compact object mergers do not necessarily take place in the host
galaxies, and may travel outside of them. With the discovery of
afterglows and identification of host galaxies one can measure
the distribution of GRBs with respect to their host galaxies.
This distribution has been calculated using different population
synthesis codes, and for different galactic potentials
\cite{1999MNRAS.305..763B,BBZ99,FWH99}. In this paper we
compare the distributions of different types of compact object
binaries: double neutron star systems (NS-NS), black hole
-neutron star systems (BH-NS) and double black holes (BH-BH).
We calculate the orbits and distributions of the projected 
distances on the sky for two extreme cases: a massive galaxy like
the Milky Way, and empty space (corresponding to e.g. a globular
cluster), and consider a wide range of possible kick velocity
distributions. We find that BH-NS are more likely gamma-ray burst
counterparts, since they lie close to the host galaxies,
contrary to the NS-NS binaries.
 } \end{abstract}


\section{Introduction}

The discovery of X-ray \cite{1997IAUC.6576....1C} and later
optical afterglows \cite{1997IAUC.6584....1G} of gamma-ray
bursts led to identification of gamma-ray burst host
galaxies \cite{1997IAUC.6588....1G}. Such host galaxies were
expected in the cosmological model, since most of the theories
identified GRBs with some, perhaps extreme, stages of stellar
evolution. However, various physical models of GRBs gave
different predictions regarding the location of GRBs with
respect to the host galaxies. In the framework of the collapsar
model, and in general all models that relate GRBs to the final
stages of evolution of massive stars \cite{Colgate}, one
expects that GRBs are found in the star forming regions. In the
second class of models where bursts are associated with mergers
of compact object binaries such associations are not obvious.
Compact object binaries may live for quite a long time before
they merge, and given the possibility that they could 
have high velocities they may travel away from the place they were
formed.

The distribution of compact object mergers can be found using
stellar population codes. The main problem of this approach is
that such codes contain a number of poorly known parameters, 
which may affect the results. One of the most important
parameter is the kick velocity a newly born compact object
receives at birth.

The distribution of double neutron star systems around galaxies
has been calculated for a few types of galaxies by Bloom et~al.
\cite*{1999MNRAS.305..763B}.  Bulik et~al.~\cite*{BBZ99}
calculated such distribution for the case of a massive galaxy
like the Milky Way, and for the case of empty space, using four
different kick velocity distributions.  These studies considered
only  binaries containing neutron stars. We used an assumption
\cite{BBZ99,BBZ99-Rome} that all supernovae lead to formation of
a  $1.4\,M_\odot$ neutron star.   However, mergers of binaries
containing a black hole are now more favored for GRBs. One
reason for this is energetics, GRB990123 had an equivalent
isotropic energy release of $10^{54}$ ergs. The energetic
requirements go down when considering that the relativistic
outflow from the central fireball is not isotropic but beamed.
On the other hand not all kinetic energy in the outflow can be
converted  into gamma rays, and therefore the energetic
requirement for the GRB central engine will go up. Therefore it
is  important to investigate the distribution around galaxies of
binaries containing black holes.

In this paper we extend the results of \cite{BBZ99}, to include 
the case of compact object binaries containing black holes.
In section 2 we describe the model for population synthesis
and galactic potential used in this paper, in section 3 
we present the results, and we summarize this work in section 4.

\section{The model}

\subsection{Population synthesis code}

We use the population synthesis code described in detail in 
Belczy{\'n}ski \& Bulik \cite*{BB1998}. Within this model we
assume that the distribution of the masses of the primary  stars
is $$\Psi(M) \propto {M}^{-1.5},$$ with $10\,M_\odot < M <
50\,M_\odot$. Such a range was chosen  because we only want to
deal with stars that may undergo a supernova explosion. The
lower limit (M$_{SN}=10\,M_\odot$) might seem high if one 
believes that large convective core overshooting is required in
order  to reproduce observations of single stars (e.g. observed
wide width of  main sequence bands). Large convective core
overshooting stellar models will result in  M$_{SN}$= 6--8
M$_\odot$. But following the arguments of \cite{VRL1998}  we
note that the  effects like stellar rotation or enhanced stellar
winds  may also explain the observations of single stars without
need  of large convective core overshooting. If one uses small
convective core overshooting then the limiting  mass of ZAMS
star to become supernova becomes larger than 9 M$_\odot$. Change
of M$_{SN}$ to a lower value would increase the number of NS-NS
binaries as compared to the number of BH binaries. This would
not change our results as in this paper we work only  on
distribution of a given class of mergers around its host
galaxy. 

The distribution of the mass ratio $q$ (secondary to primary
mass) is 
$$\Phi(q) \propto {\mathrm {const}}$$
 with $0<q<1$.
The distribution of the initial binary eccentricity $e$ is
$\Xi(e) = 2e$ with $0<e<1$, and 
the distribution of the initial semi-major axis
$a$ used in population synthesis codes is 
flat in the logarithm, i.e.
$\Gamma(a) \propto {a}^{-1} $ with the maximum $a_{max}=10^5
R_\odot$.

Very little is known observationally about the kicks newly born
black holes receive. While studies of pulsars  indicate that
substantial kicks are possible  for newly born neutron stars, we
believe that black hole kicks  are smaller. The physical reason
could be just that black holes are more massive, and also that
asymmetric neutrino emission  may not take place when a black
hole is formed. Lipunov et~al. \cite*{1997MNRAS.288..245L} use a
Gaussian parameterization of the kick velocity and assume that
for black holes it is proportional to the mass lost in the
explosion. Fryer et~al.~\cite*{FWH99} show results for several
cases, using the same functional form of the kick  distribution 
for black holes and neutron star and  assuming that black hole
kicks are ten times smaller than those for neutron stars.

We parameterize the distribution of the kick velocity a newly 
born compact object receives in a supernova explosion by a
three  dimensional Gaussian with the width $\sigma_v$, and
consider several values of this parameter. Such a choice allows
to compare the distributions of  binaries containing black holes
vs. the double neutron stars that could be formed with different
kick velocities.

For a  detailed description of the population synthesis code 
we refer the reader to \cite{BB1998}. There is only one
modification to the code, i.e. in the choice of the mass of a
compact object formed in a supernova explosion. This is described
in the next subsection.

\begin{figure}

\includegraphics[width=\columnwidth]{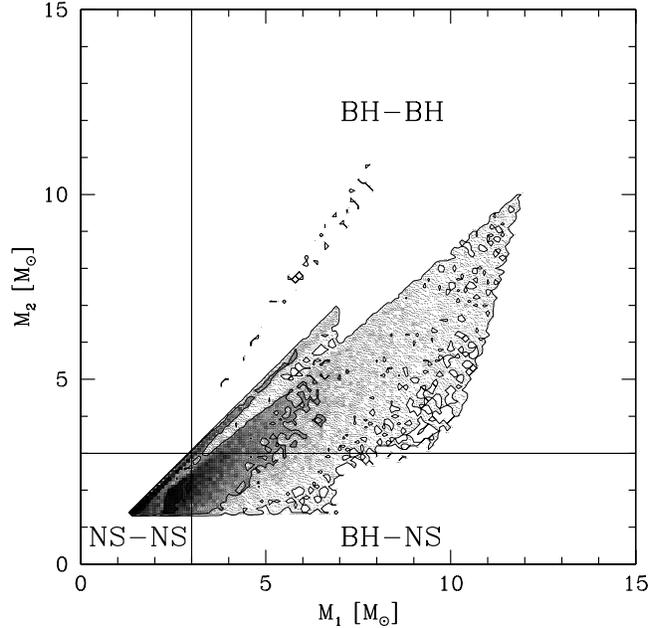}

\caption{ The population of the compact object binaries in the
space spanned by their masses. The three regions correspond to
different initial mass ratios: the binaries with $M_1$ reaching
up to $10\,M_\odot$ and  $M_2< M_1$, originate from  systems
with the initial mass ratio $q<0.88$,  and the binaries along
the line $M_1\approx M_2$ originally had nearly equal masses
$q>0.95$. The sparse population of systems for which $M_2 > M_1$
originates from binaries with $0.88<q<0.95$. } 
\label{masybw}
\end{figure}

\subsection{Compact object formation in a Sn explosion}

In general, supernovae from massive stars should lead to
formation of more massive objects, perhaps black holes, and
supernovae from less massive stars to lighter compact objects
like neutron stars. The boundary between these two possibilities
and the mass of a compact object formed in an explosion is not
clear. Lipunov et~al.~\cite*{1997MNRAS.288..245L} assumed that a
neutron star with a mass $1.4\,M_\odot$ is formed for
progenitors with core masses below  $35\,M_\odot$ at the time of
explosion, and for larger masses  black holes are formed with
the mass $k_{bh}\times 35\,M_\odot$ where $k_{bh}=0.3$. The
parameter $k_{bh}$ is varied between $0.1$ and $1$
\cite{Lipunov-Kyoto}. Tout et~al.~\cite*{1997MNRAS.291..732T}
used a different prescription: neutron stars are born with
masses equal to $M_{Ch} + M_{pro}/50$, where $M_{Ch}=
1.4\,M_\odot$ is the Chandrasekhar mass, and $M_{pro}$ is the
progenitor mass at the time of collapse, and assume that object
above $1.8\,M_\odot$ are black holes. Fryer
et~al.~\cite*{Fryer99} presented numerical simulations of core
collapse in massive stars, and found that neutron stars are
formed for progenitors with initial masses  below $22\,M_\odot$,
and black holes for masses above this value. In a population
synthesis code \cite{FWH99} assume simply that a compact object
with a mass ${1\over 3}M_{p}$, where $M_{p}$ is the progenitor
mass at the time of collapse. On the other hand Ergma and van
den Heuvel \cite*{1998A&A...331L..29E} analyze the population of
known X-ray binaries and argue that the initial mass of the
progenitor  is not the sole parameter determining the mass of a
compact object formed in a supernova explosion, but probably
other  factors like rotation or the magnetic field may play an
important role. 

Given such a range of uncertainties we use the following
prescription  for the compact object mass (either neutron star
or black hole): the compact object mass is equal to half of the
mass of the final helium core of progenitor star. This approach
gives higher masses of compact objects as compared to the 
prescriptions discussed above. However, simulations of core
collapse during supernova  explosions \cite{FWH99}  show that
the compact object might accrete back some of the  material
firstly ejected in explosion for a fraction of the range of
stellar masses we are concerned in our work.
Although the predictions how much mass is accreted in a given
case are still uncertain and can not be easily parameterized but
we believe  that  some enhancement of compact object mass is
reasonable.

 Standard equations of state for neutron stars give the maximum
mass of neutron star to be between 1.9 and 2.6 M$_\odot$,
although the rotation  may increase it by about another 0.2
M$_\odot$.  We have used value of 2.4 M$_sun$ as the maximum
neutron star mass in our previous  work \cite{BBZ99}.  However,
there are equations of state for neutron stars which go as high 
as 3 M$_\odot$ or even up to 3.2 M$_\odot$ if rotation is
included \cite{1994ApJ...424..823C}. Thus in this work we assume
that the  maximum  neutron star mass is  3.0 M$_\odot$. This
number was not crucial in the previous paper, as we treated
both  groups, namely double neutron stars and black hole neutron
stars binaries  together and in \cite{BBZ99} we assumed that all  
supernovae lead to formation of a $1.4\,M_\odot$ neutron star.
In present work this number plays an important role  and as we
are mostly interested in black hole binaries we want to be  sure
that we do not include any neutron stars as black holes in the 
present calculations. Thus we  use 3.0 M$_\odot$ as the upper
limit on neutron star mass and consequently the  dividing line
between neutron star and black hole.

\begin{figure*}

\includegraphics[width=\textwidth]{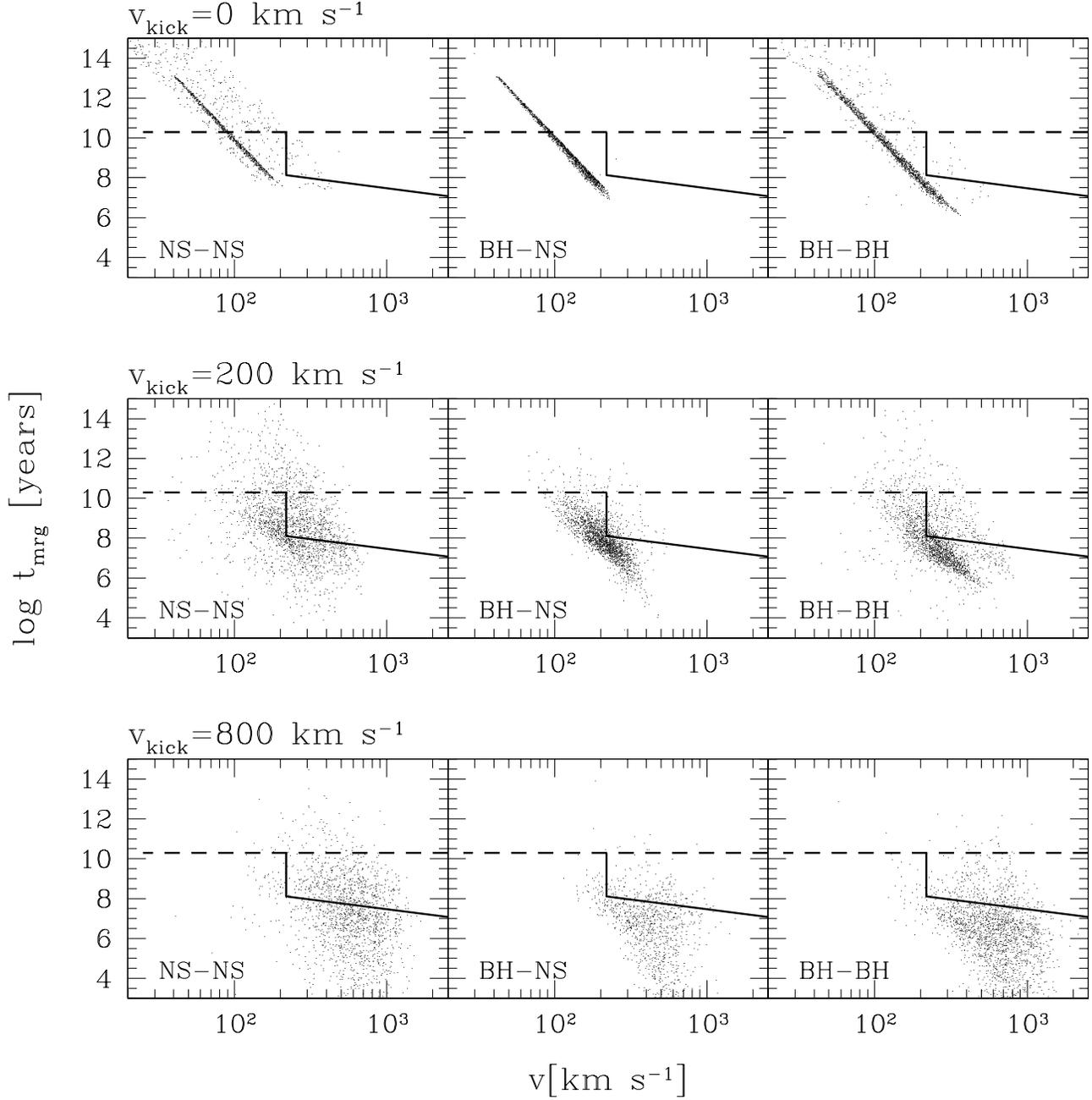}

\caption{The distributions of binaries in the plane spanned by 
the center of mass velocity $v$ and merger time $t_{mrg}$.
In each panel we also plot the following lines:
the horizontal dashed line
corresponding to the Hubble time (15Myrs);
two solid lines: the vertical
corresponding to $v=200\,$km~s$^{-1}$ - a lower limit on the
typical escape
velocity from a galaxy, and the line corresponding to a constant
value of
$v\times t_{merge} =30\,$kpc. The panels correspond to the kick
velocity distributions width of $0$km~s$^{-1}$, $200$km~s$^{-1}$,
and $800$km~s$^{-1}$ from top to bottom respectively. In each
panel we present the distribution of NS-NS, BH-NS and BH-BH
separately.}
\label{vtpap}
\end{figure*}

\begin{figure*}

\begin{center}
\vspace{-2mm}
\begin{tabular}{ll}
$\sigma_v = 0\,$km~s$^{-1}$; Empty space &
$\sigma_v = 0\,$km~s$^{-1}$; Galaxy\\
\includegraphics[width=0.38\textwidth]{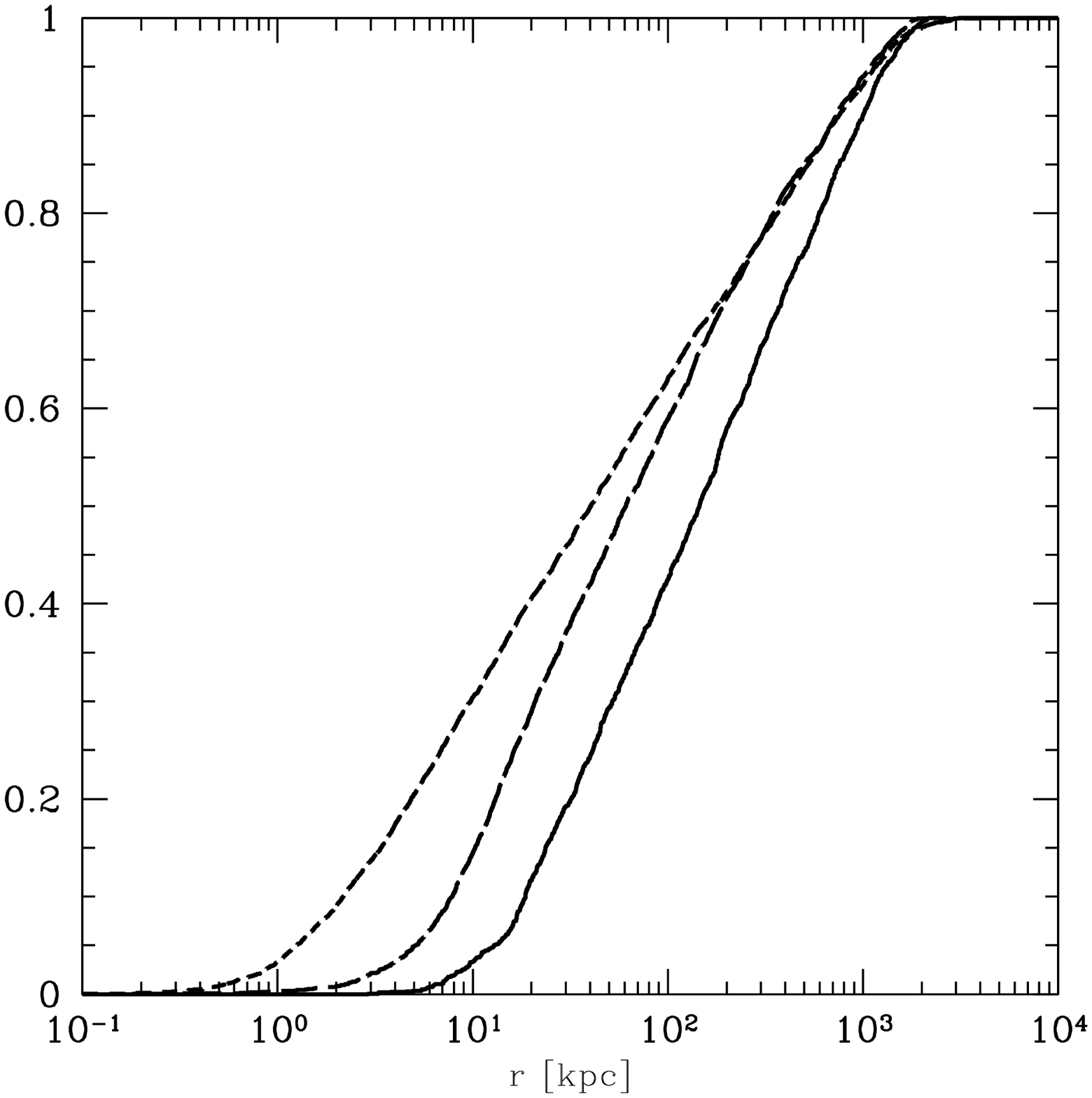} &
\includegraphics[width=0.38\textwidth]{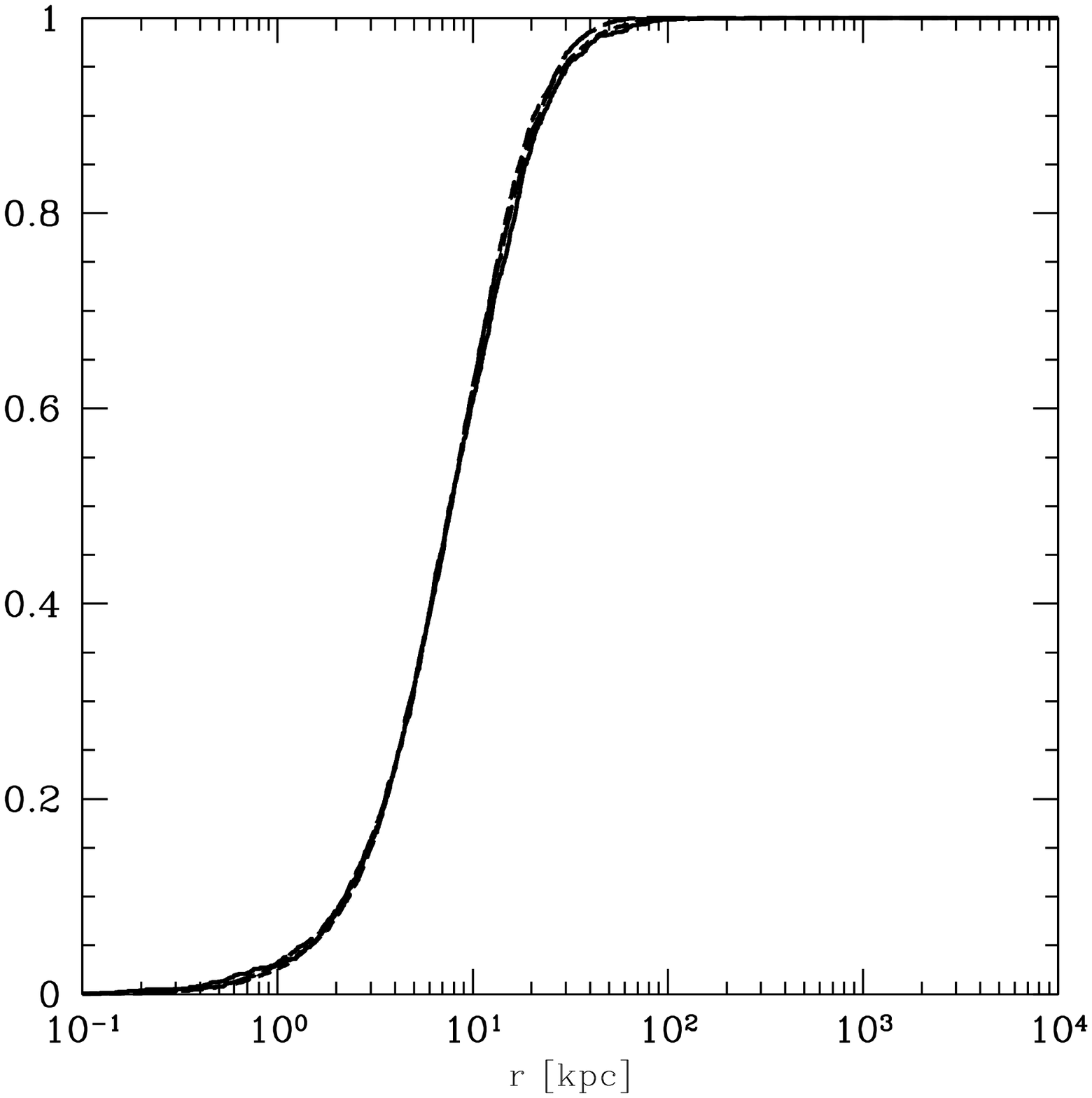} \\
$\sigma_v = 200\,$km~s$^{-1}$; Empty space &
$\sigma_v = 200\,$km~s$^{-1}$; Galaxy\\
\includegraphics[width=0.38\textwidth]{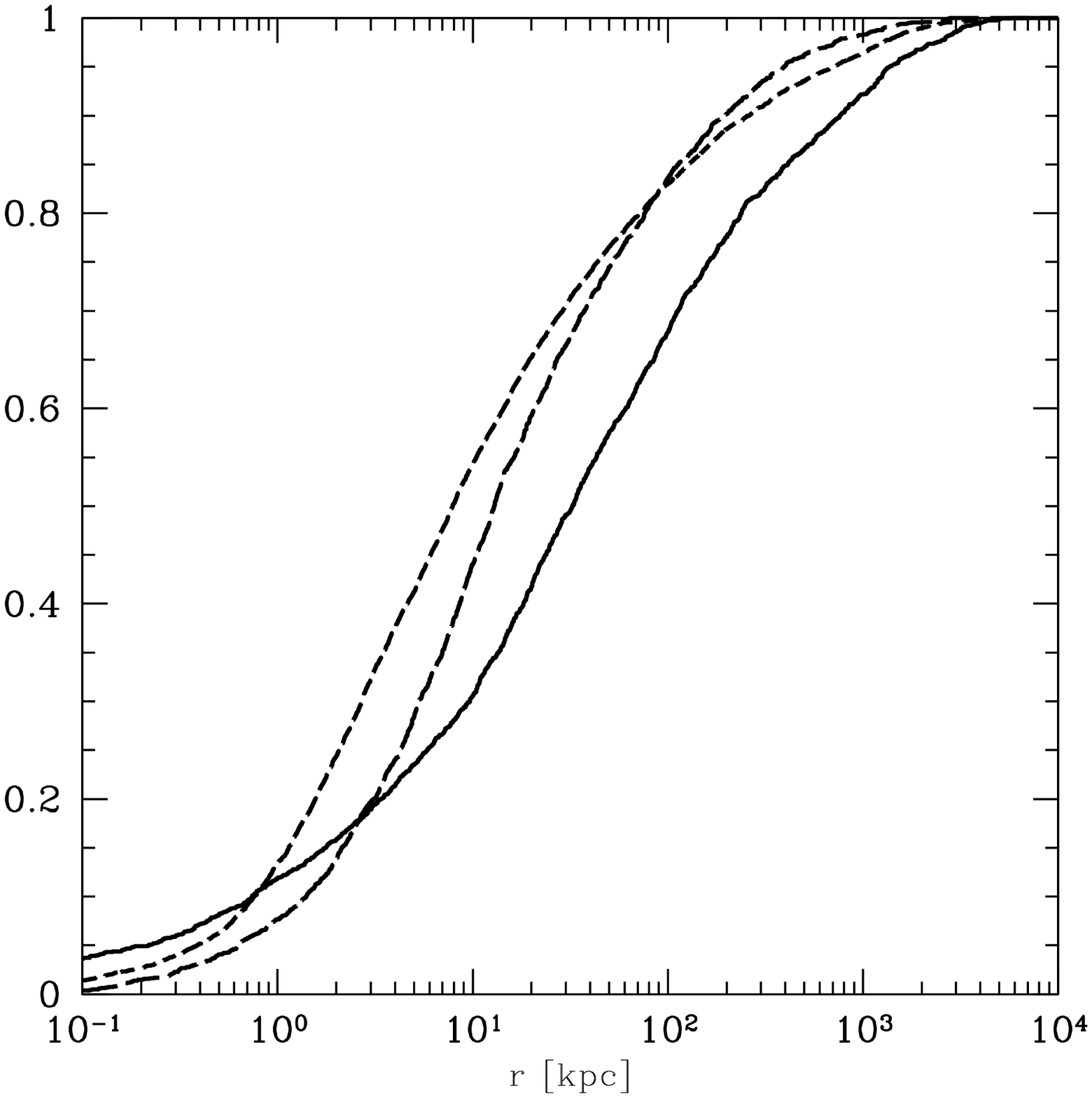} &
\includegraphics[width=0.38\textwidth]{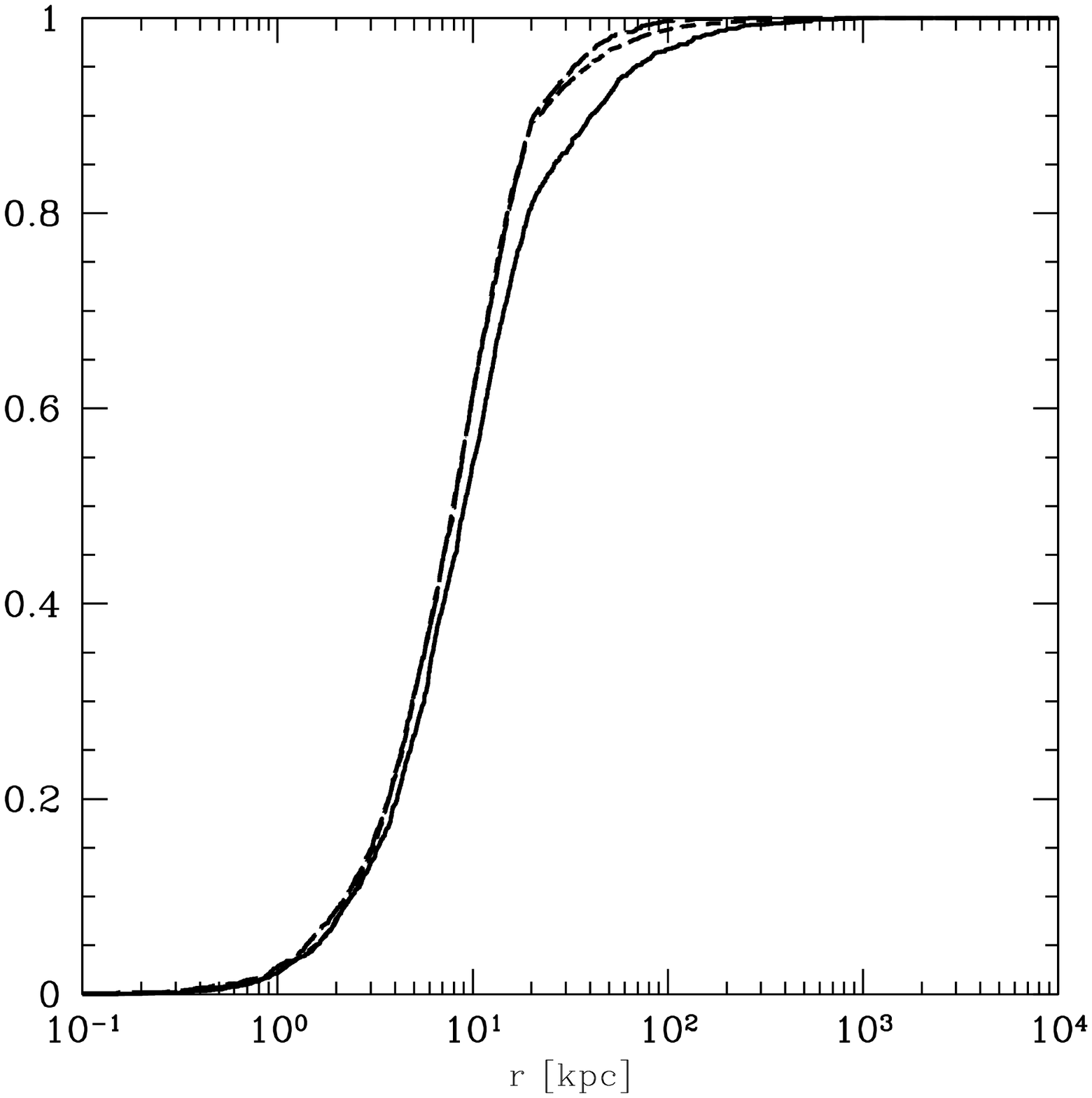} \\
$\sigma_v = 800\,$km~s$^{-1}$; Empty space &
$\sigma_v = 800\,$km~s$^{-1}$; Galaxy\\
\includegraphics[width=0.38\textwidth]{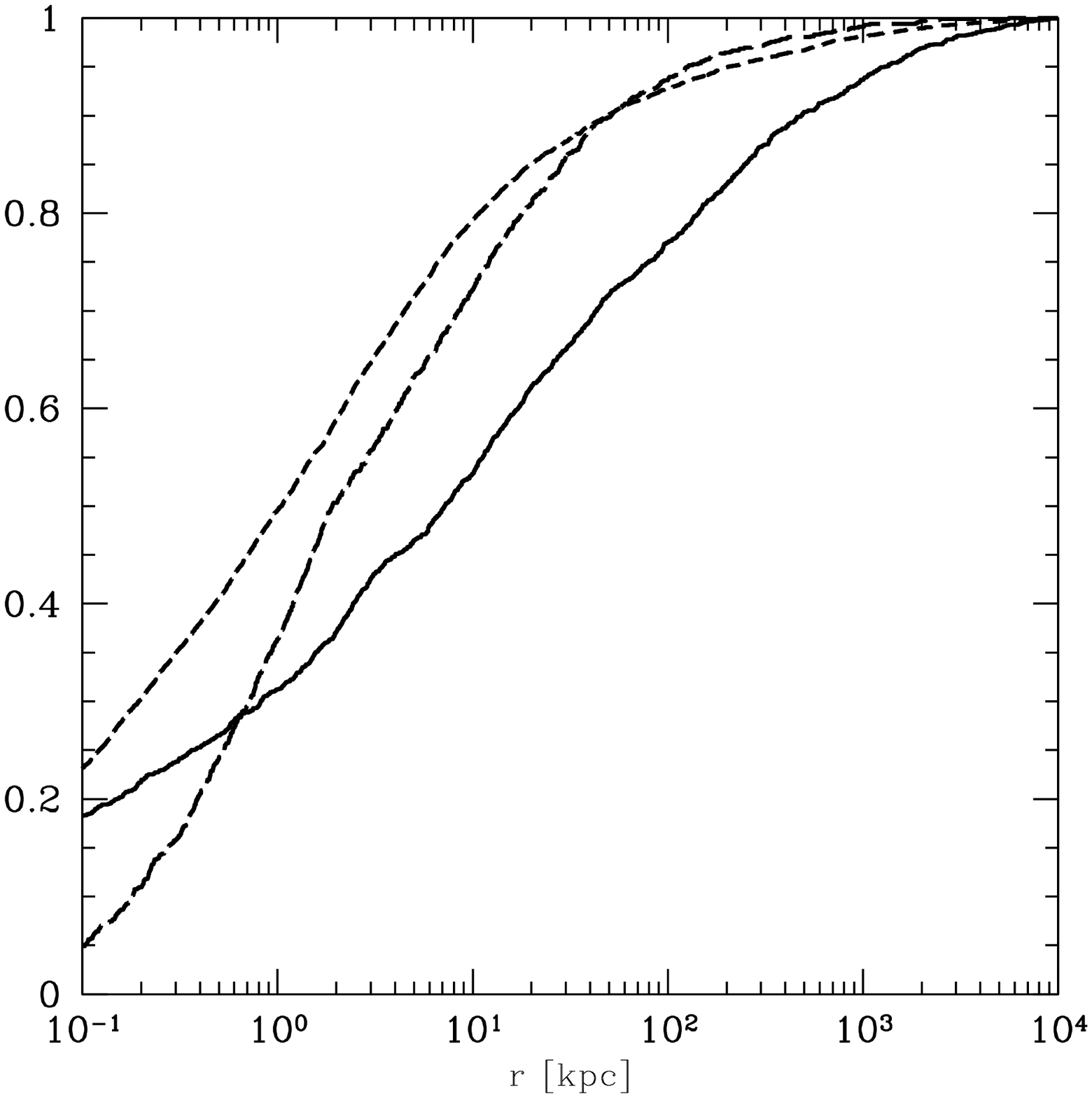} &
\includegraphics[width=0.38\textwidth]{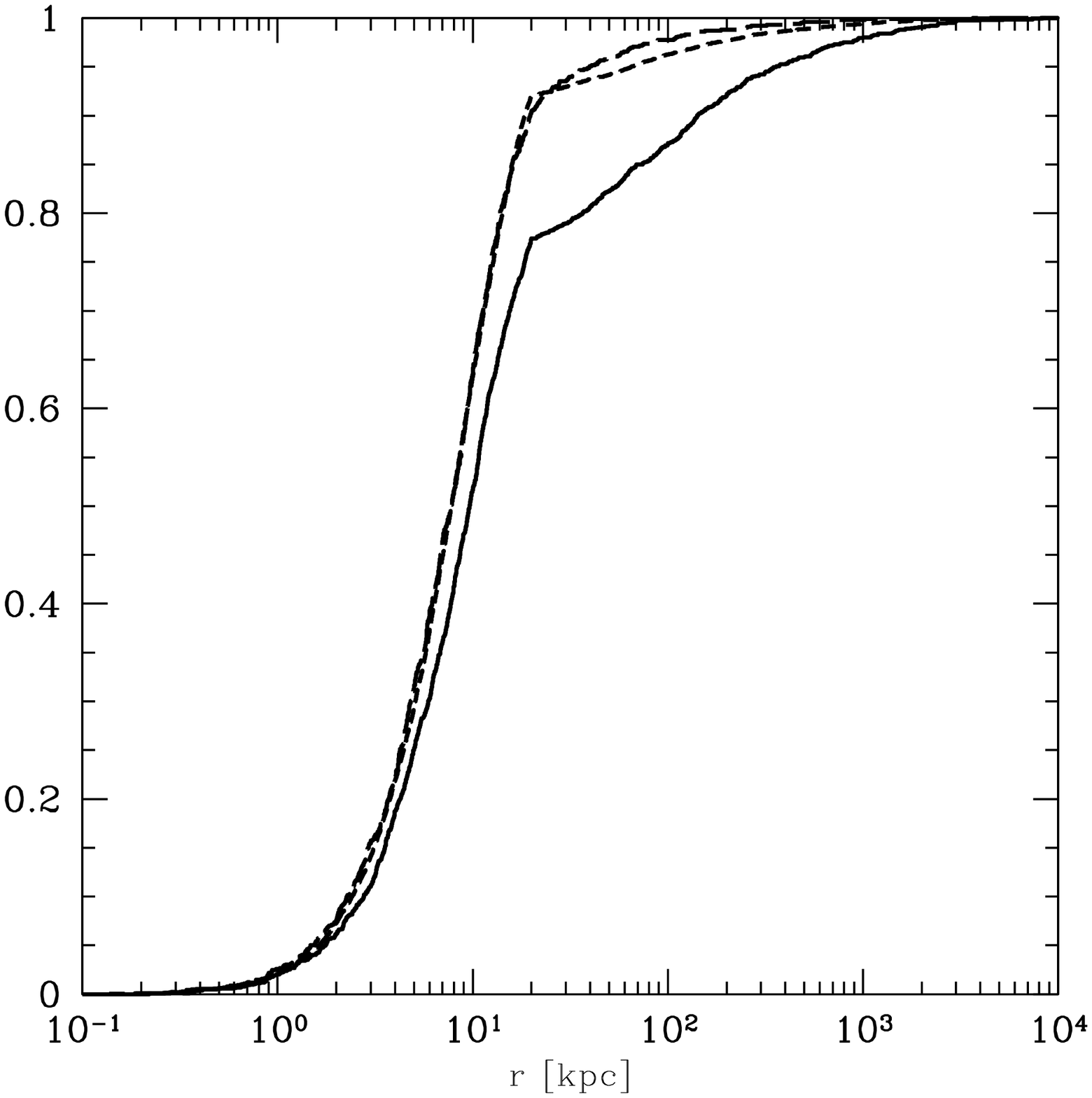} 
\end{tabular}
\end{center}
\vspace{-6mm}
\caption{Cumulative distribution of projected distances on the
sky of compact object binary mergers. The left panels
correspond  to the case   of propagation in the empty space, and
the right panels  to the propagation in the potential
of a massive galaxy. The top panels show the case of
no kick velocities $\sigma_v=0\,$km\,s$^{-1}$, the middle panels
correspond to $\sigma_v=200\,$km\,s$^{-1}$, and the lower
panels represent the case $\sigma_v=800\,$km\,s$^{-1}$.
In each panel the solid line, long dashed , and short dashed
lines represent the the distributions of NS-NS, BH-NS, and BH-BH
binaries, respectively.}
\label{lv}
\end{figure*}

\subsection{Gravitational potentials}

We follow the approach we have used in our previous paper
\cite{BBZ99}, and consider two extreme cases: propagation in a
potential of a massive galaxy like the Milky Way, and
propagation in empty space (corresponding to e.g. a globular
cluster origin). The potential of a massive galaxy  consists of
three components: bulge, disk, and halo.  To model the bulge and
disk potentials we use the potential model proposed by
\cite{1975PASJ...27..533M}: \begin{equation} \Phi(R,z) =
{GM_i\over \sqrt{ R^2 + (a_i + \sqrt{ z^2 + b_i^2})^2}} 
\end{equation} where $a_i$ and $b_i$ are the parameters, $M$ is
the mass, and $R=\sqrt{x^2 +y^2}$.  The dark matter  halo
potential is spherically symmetric \[ \Phi(r) = - {G M_h\over
r_c} \left[ {1\over 2} \ln\left( 1 + {r^2\over r_c^2}\right) +
{r_c\over r} {\mathrm{atan}} \left( r\over r_c\right) \right] \]
where $r_c$ is the core radius. The halo potential corresponds
to a mass distribution $\rho = \rho_c/[1 + (r/r_c)^2]$, and we
introduce a cutoff radius $r_{cut}=100\,$kpc beyond which the
halo density falls to zero, in order to make the halo mass
finite and the halo gravitational potential  is $\Phi(r)\propto
r^{-1}$ when $r>r_{cut}$. We use the following values of the
parameters derived for the Milky Way (we assume that the Milky
Way is a good example of a massive  galaxy;  the bulge potential
($i=1$): $a_1 =0\,$kpc, $b_1 = 0.277\,$kpc, $M_1 = 1.12\times 10
^{10}\,M_\odot$; the disk potential ($i=2$): $a_2 = 4.2\,$kpc,
$b_2 = 0.198\,$kpc,  $M_2 = 8.78\times 10^{10}\,M_\odot$; the
halo potential: $r_c =
6.0\,$kpc, and $M_h = 5.0 \times 10^{10}\,M_\odot$
\cite{1990ApJ...348..485P,1991ApJ...381..210B}. The distribution
 of stellar initial positions in the model galaxy is a double
exponential $ P(R,z) = R \exp(-R/R_{exp}) \exp(-z/z_{exp})$,
with $R_{exp}= 4.5\,$kpc, $z_{exp}=75\,$pc, and we cut the
distribution at $R_{max}= 20\,$kpc \cite{1998ApJ...505..666B}. 
The initial velocity of a binary is assumed to be equal to the
local  rotation velocity in the galactic disk. After each
supernova explosion we add the velocity the system received,
which is calculated in the population synthesis code. We follow
the trajectories of the binaries until they merge.

\section{Results}

In order to see the general properties of the population of
compact object binaries we consider the case with no kick 
velocities $\sigma_v=0\,$km\,s$^{-1}$. In Figure~\ref{masybw} we
present the density of binaries as a function of their masses.
The population in Figure~\ref{masybw} has three components. 
The main component are the binaries which originate from systems 
with small mass ratios $q<0.88$, the binaries along the line 
$M_1\approx M_2$ originally had nearly equal masses $q>0.95$ and 
the population of systems for which $M_2 > M_1$ originates from 
binaries with intermediate mass ratios $0.88<q<0.95$. 
These components come from different regimes of initial binary mass 
ratio, which to some degree sets the subsequent binary evolution, as
discussed earlier by \cite{Bethe1998}. 
The detailed description of evolution in each of these regimes can 
be found in Belczy{\'n}ski \& Bulik \cite*{BB1998}. 

When considering higher kick velocities the basic shape of  the
diagram of Figure~\ref{masybw} does not change substantially. We
draw lines corresponding to $M_1 = 3M_\odot$ and $M_2 =
3M_\odot$,  the upper limit on the maximal
mass of a neutron star. We predict a large population of compact
object binaries that contain a black hole.  While we do not
expect to see double black hole binaries, it is  puzzling that
no black hole neutron star binaries are known among  binary
pulsars.  These could be explained by the fact that so far we
know only a  few of such objects. 

On the other hand one can argue that before the second supernova
explosion the stars are tidally locked, which would prevent the
newly born neutron star from becoming a pulsar, provided that
the orbit before the explosion is wide enough. There are two
effects bracketing  the size of orbit of binaries before second
supernova explosion, and  they have chance (depending on
evolutionary parameters) to select  binaries, which are tidaly
locked before second supernova explosion  and  are not too tight
at that time. The first selection occurs during the first
supernova explosion which tends to  disrupt the widest binaries,
as their binding energy is small. After the first supernova
explosion the population of binaries is  reduced to these which
are close, with relatively small orbits, and which  are in
general highly eccentric (the kick has random direction). As the
system evolves toward second supernova, the secondary star
enters consecutive giant stages and there is a good chance that
it will fill a significant part of its Roche lobe and become
tidaly synchronized. On the other hand the system can not be too
tight to allow for the secondary giant evolution and thus
minimum separation is set by the size of evolved secondary. 

There could also be a gap in the distribution of compact object
formed in supernova explosion, because of the fall-back of matter 
onto a compact object in the supernova explosion \cite{Fryer99}.
If such a gap really exists than the number of BH-NS binaries is
smaller than our simulations predict.

The properties of the population of compact object binaries
can be found from Figure~\ref{vtpap} where we plot the
distribution of different types of objects in the plane:
$v$ -the center of mass velocity, and $t_{mrg}$ - time to merge.
In the the top panel, corresponding to the case
of no kick velocities, $\sigma_v = 0\,$km\,s$^{-1}$, a 
correlation between $v$ and $t_{mrg}$ is apparent.
The center of mass velocities in this case are due to mass
ejection from the system \cite{1960BAN....15..265B}, and is
of the same order of magnitude as the orbital velocity.
The lifetime due to gravitational wave energy loss has been 
calculated by Peters~\cite*{Peters1964}, and it scales
like $t_{mrg} \propto a^4$, where $a$ is the orbital separation.
The orbital separation is inversely proportional to the square of
the orbital velocity, hence $t_{mrg} \propto v^{-8}$, which
explains the trend in the top panel of Figure~\ref{vtpap}.
With increasing the kick velocity only very tight and eccentric
systems survive and their lifetime becomes shorter.
The typical center of mass velocity is now determined by 
both the orbital velocity at the time of supernova explosion, and
the kick velocity. This moves the center of the distribution
towards shorter merger times and higher velocities with
increasing the kick velocity. The distributions become  wider
and the correlation between $v$ and $t_{mrg}$ vanishes for the case
of high kick velocities.

Similarly to our previous work \cite{BBZ99}  we present the
distribution of  compact object mergers projected, distances on
the sky for two  cases: (i) propagation in the potential of a
massive galaxy like the Milky Way, and (ii) propagation in empty
space, corresponding to  a small galaxy. The results  the case
(i) and (ii) are shown in the right and left panels of
Figure~\ref{lv}, respectively. Generally for the same width of
the kick velocity distribution, the spatial distribution of more
massive binaries  is tighter around the origin for the case
(ii). This is due to the fact that the merger time for the more
massive  objects is typically smaller, see also
Figure~\ref{vtpap}. The median projected distance for the BH-BH
binaries is  about ten times smaller than that for the NS-NS
binaries. Increasing the kick velocity leads to two effects:
decrease of the lifetime of a compact object binary and increase
of its center of mass velocity. The decrease of the lifetime,
however, is more important in determining the spatial
distribution of these objects. The median distance to the
merger  decreases significantly with increasing the kick
velocity, however,  a long tail of the distribution extending to
large distances remains.

In the case(i) (propagation in the potential of a massive
Galaxy) the increase of the width of the kick velocity
distribution leads to division of the population into two
groups: the bound and the escaping binaries. The spatial
distribution  of bound binaries follows that of the stellar
matter in galaxy. This division is responsible for the breaks in
the cumulative plots in the right panels of Figure~\ref{lv}. 
There is a  small difference between the populations of
different types of compact object binaries, regardless of the
kick velocity. With increasing the kick velocity the fraction of
unbound binaries becomes larger. This fraction also depends on
the type of binary, the escaping fraction of binaries containing
black holes is smaller than that of the NS-NS binaries. For the
case of the highest kick velocity distribution width considered
here, $\sigma_v=800\,$km\,s$^{-1}$ less than 10 percent of black
hole neutron star binaries become unbound in comparison to
almost 30 percent of the double neutron star systems.

\section{Discussion}

We have presented an extension of our previous work \cite{BBZ99}
where we analyzed the distribution of neutron star binaries
around galaxies. Here we have modified the code to produce 
binaries which are heavier, e.g. contain black holes. The class
defined as NS-NS binaries in this paper is broader than the
binaries in \cite{BBZ99}, since it includes objects with masses
up to $3\,M_\odot$, while the heaviest compact objects in our
previous paper had a mass of $2.4\,M_\odot$, and the bulk of
them were $1.4\,M_\odot$ neutron stars. The trend, which we also
noted before, that heavier binaries tend to  stick closer to the
host galaxies is evident when comparing the distribution around
a massive galaxy for the NS-NS class shown in this paper with
the distributions shown in \cite{BBZ99}. For the highest kick
velocity distribution width considered here,
$\sigma_v=800\,$km\,s$^{-1}$, the fraction of escaping NS-NS
binaries is $\approx 25$\% here, in comparison to almost $40$\%
in \cite{BBZ99}.

When interpreting the results of Figure~\ref{lv} one has to 
remember that for reasons stated above the kick velocity 
probably decreases with increasing mass of the compact object.
Thus, the distribution of NS-NS binaries lies somewhere between
the case $\sigma_v = 200\,$km\,s$^{-1}$ and 
$\sigma_v = 800\,$km\,s$^{-1}$ (the middle and the lower
panel in Figure~\ref{lv}, and the distribution of 
BH-NS binaries must be somewhere between the cases
$\sigma_v = 0\,$km\,s$^{-1}$ and $\sigma_v = 200\,$km\,s$^{-1}$.
We find that BH-NS binaries (and also BH-BH systems) will
merge within the host galaxies provided that they are massive.
In the case of smaller galaxies, the BH-NS mergers
will take place typically a few times closer to the 
host galaxy than the NS-NS mergers. However, there is
still a long tail of the distribution, with a
significant fraction extending to large distances.

We have shown that the distances from the host galaxies where
compact object binaries merge decrease with increasing the mass
of the binary. The reasons for such a behavior are twofold.
First, the lifetimes of heavier binaries are smaller, because of
the increased gravitational wave energy loss. Second, the width 
of the kick velocity distribution in a supernova explosion is
probably smaller when black holes are formed, which leads to
smaller center of mass velocities.  We conclude that BH-NS
binary mergers are more likely progenitors of gamma-ray bursts
than the NS-NS binaries. Their distribution closely follows that
of the matter in massive galaxies, and also allows to explain
the extreme energetics of some bursts. However, this conclusion
is so far  based  on a small sample of well observed GRB
afterglows  for long and hard bursts.

\acknowledgements
This work has been supported by the KBN grants
2P03D01616, and 2P03D00415 and also made use of the NASA
Astrophysics Data System.

\bibliography{aamnem99,Mergers}
\bibliographystyle{aabib99}

\end{document}